\documentclass[12pt,a4paper]{article} 
\usepackage[latin1]{inputenc}
\usepackage[T1]{fontenc} 
\usepackage[english]{babel}
\usepackage{amsmath,amssymb,amsfonts} 
\title{Casimir interaction and gauge invariance in finite width mirrors} 
\author{C.~D.~Fosco
and
M.~L.~Remaggi\\
{\normalsize\it Instituto Balseiro}\\ 
{\normalsize\it Universidad Nacional de Cuyo}\\ 
{\normalsize\it Centro At\'omico Bariloche}\\ 
{\normalsize\it Comisi\'on Nacional de Energ\'\i a At\'omica}\\ 
{\normalsize\it R8402AGP Bariloche, Argentina}} 
\begin{document} 
\date{} 
\maketitle 
%====================================================================
\begin{abstract} 
	We study the general form of the Casimir interaction between two
	flat	finite-width, parallel, infinite mirrors, described by
	their respective vacuum polarization tensors, under the assumption
	of translation invariance along the directions parallel to the
	mirrors' planes.  Their properties along the normal coordinates are
	left, on the other hand, rather arbitrary, but constrained  by the
	Ward identity for the vacuum polarization tensors. We construct
	solutions to those identities, evaluating the corresponding Casimir
	energy of the system by reducing the problem to a collection of
	one-dimensional systems.
\end{abstract}
%%%%%%%%%%%%%%%%%%%%%%%%%%%%%%%%%%%%%%%%%%%%%%%%%%%%%%%%%%%%%%%%%%%%%%%%%%
%%%%%%%%%%%%%%%%%%%%%%%%%%%%%%%%%%%%%%%%%%%%%%%%%%%%%%%%%%%%%%%%%%%%%%%%%%
%%%%%%%%%%%%%%%%%%%%%%%%%%%%%%%%%%%%%%%%%%%%%%%%%%%%%%%%%%%%%%%%%%%%%%%%%%
%%%%%%%%%%%%%%%%%%%%%%%%%%%%% Introduction %%%%%%%%%%%%%%%%%%%%%%%%%%%%%%%
%%%%%%%%%%%%%%%%%%%%%%%%%%%%%%%%%%%%%%%%%%%%%%%%%%%%%%%%%%%%%%%%%%%%%%%%%%
%%%%%%%%%%%%%%%%%%%%%%%%%%%%%%%%%%%%%%%%%%%%%%%%%%%%%%%%%%%%%%%%%%%%%%%%%%
%%%%%%%%%%%%%%%%%%%%%%%%%%%%%%%%%%%%%%%%%%%%%%%%%%%%%%%%%%%%%%%%%%%%%%%%%%
\section{Introduction}\label{sec:intro} 
In recent years, it has become increasingly clear that the use of sensible
models for the description of the electromagnetic (EM) properties of the
media that constitute the mirrors is an important step for the improvement
of Casimir energy calculations~\cite{Casimir,rev}.  The most relevant of those
properties may be accounted for in different ways, for instance, in an
effective theory for the EM field, obtained after the functional integration of the
matter degrees of freedom inside a mirror~\cite{Fosco:2008td}.  

In this letter, we shall consider two infinite, parallel, finite-width
mirrors, the properties of which, in a Casimir energy calculation, are
described by their vacuum-polarization tensors.  Those tensors
reflect the assumption that the model is translation invariant along
the spatial directions $(x_1,\, x_2)$, parallel to the (constant-$x_3$)
planes of the mirrors, as well as time-independent. We summarize this
symmetry by saying that the model is invariant under translations in
\mbox{$x_\parallel \equiv (x_0,x_1,x_2)$}, where $x_0$ denotes the
Euclidean time coordinate.  

Regarding $x_3$, the {\em normal\/} direction, the very presence of the
mirrors breaks translation symmetry of the system along that coordinate.
Because of this (specially for finite-width mirrors) there are different
possibilities regarding their structure along $x_3$.  There are, however,
non trivial constraints on that dependence, that follow from the Ward
identity for the vacuum polarization tensors, as they entangle the
$x_\parallel$ and $x_3$ dependencies.  The solution to those constraints is
not unique. It is worth emphasizing that finite width mirrors coupled to
the EM field have been considered in the literature~\cite{rev}, but usually their
properties along the normal direction are only functions of the frequency
of the incident waves.

In this letter, we construct non trivial solutions to those constraints,
obtained by using some extra simplifying assumptions, and evaluate
${\mathcal E}$, the vacuum energy per unit area.  Regarding the way to
introduce that observable, we shall write ${\mathcal E}$ as the
in terms of an effective action $\Gamma$, which can in turn
be expressed as a function of the vacuum transition amplitude ${\mathcal Z}$:
\begin{eqnarray}\label{eq:defedens}
	{\mathcal E}&=&\lim_{T \to \infty} ( \frac{\Gamma}{T} )
	\nonumber\\ 
	\Gamma &=&-\lim_{L_\parallel \to \infty}
	\Big(\frac{1}{L_\parallel^2}\,\log\frac{{\mathcal
	Z}}{{\mathcal Z}_0}\Big) \;,
\end{eqnarray}
where $L_\parallel$ is a length that characterizes the (parallel) size of
the plates, $T$ denotes the extent of the time coordinate, and ${\mathcal
Z}$ is the vacuum transition amplitude
for (square) plates of length $L_\parallel$.

${\mathcal Z}$ may then be expressed as an Euclidean functional
integral:
\begin{equation}\label{eq:defzetabeta}
{\mathcal Z} \;=\; \int \big[{\mathcal D}A\big] \, e^{-{\mathcal S}(A)} \;,
\end{equation}
where ${\mathcal S}(A)$ denotes the Euclidean action for the gauge field
$A$ in $3+1$ dimensions:
\begin{equation}\label{eq:defsa}
{\mathcal S}(A) \;=\; {\mathcal S}_0(A) \,+\,{\mathcal S}_I(A) \;,
\end{equation}
${\mathcal S}_0$ is the action for the free EM field:
\begin{equation}
	{\mathcal S}_0(A) \;=\; \frac{1}{4} \, \int d^4x \, F_{\mu\nu}
	F_{\mu\nu} \;\;,\;\;\; F_{\mu\nu} = \partial_\mu A_\nu -
	\partial_\nu A_\mu \;,
\end{equation}
while ${\mathcal S}_I$ describes the coupling between $A$ and the two
mirrors. 

The parallel spatial coordinates are assumed to
be confined to a (square) box of side length $L_\parallel$, with Dirichlet
boundary conditions on the box faces~\footnote{The final result, for
$L_\parallel \to \infty$, shall be insensitive to the choice of boundary
conditions on that spatial box.}.  The path integration measure
$\big[{\mathcal D}A\big]$ is assumed to include a gauge fixing, which we
shall specify when actually performing the calculation of the integral (to
render it in the simplest possible form).

This letter is organized as follows: in section~\ref{sec:themodel} we
define the general structure of the models we consider, in terms of
the coupling between each mirror and the gauge field, 
given by localized vacuum polarization tensors. 
We then use the Ward identity for those tensors, a step which reduces the number of
independent components. 
In section~\ref{sec:reduction}, we introduce the gauge-fixing and reduce the
problem to one dimensional systems. In~\ref{sec:energy} we obtain a
compact expression for ${\mathcal E}$ in terms of the independent invariants 
characterizing each mirror. In particular, for the case of response functions which are
local in $x_3$, we obtain an even more compact expression in terms of a Lifshitz formula.  
In~\ref{sec:conclusions} we present our conclusions.
%%%%%%%%%%%%%%%%%%%%%%%%%%%%%%%%%%%%%%%%%%%%%%%%%%%%%%%%%%%%%%%%%%%%%%%%%%%%%%%
%%%%%%%%%%%%%%%%%%%%%%%%%%%%%%%%%%%%%%%%%%%%%%%%%%%%%%%%%%%%%%%%%%%%%%%%%%%%%%%
%%%%%%%%%%%%%%%%%%%%%%%%%%%%%%%  The model %%%%%%%%%%%%%%%%%%%%%%%%%%%%%%%%%%%%
%%%%%%%%%%%%%%%%%%%%%%%%%%%%%%%%%%%%%%%%%%%%%%%%%%%%%%%%%%%%%%%%%%%%%%%%%%%%%%%
%%%%%%%%%%%%%%%%%%%%%%%%%%%%%%%%%%%%%%%%%%%%%%%%%%%%%%%%%%%%%%%%%%%%%%%%%%%%%%%
\section{The model}\label{sec:themodel}
The EM field is coupled to two finite-width mirrors, denoted by $L$ and
$R$, located~\footnote{The location is defined by the $x_3$ coordinate
corresponding to their midpoints.} at $x_3=0$ and $x_3=a$, respectively.  
The model is thus specified by defining the interaction term, ${\mathcal
S}_I$, appearing in (\ref{eq:defsa}). This, in turn, consists of two terms, each one 
describing the coupling between $A$ and a single mirror:
\begin{equation}
	{\mathcal S}_I(A) \;=\; {\mathcal S}_L(A) \,+\, {\mathcal S}_R(A)
	\;.
\end{equation}
Since the forthcoming analysis is essentially the same for each mirror, to
construct the ${\mathcal S}_L$ and ${\mathcal S}_R$ terms, we first study
the form of the interaction term, which we denote by ${\mathcal S}_M$,
corresponding to a single mirror centered at $x_3=0$. Then, we shall define
${\mathcal S}_L$  and ${\mathcal S}_R$ in terms of ${\mathcal S}_M$, just by
using the respective tensor.

The ${\mathcal S}_M(A)$ term shall have the following structure:
\begin{equation}
{\mathcal S}_M (A) \;=\; \frac{1}{2} \int_{x,y} A_\mu(x)
\Pi_{\mu\nu}(x;y) A_\nu(y) \;,
\end{equation}
where we have assumed linear response for the media, $\Pi_{\mu\nu}(x;y)$ is
the vacuum polarization tensor for the mirror, and we have used a shorthand
notation for the integral over space time coordinates
\mbox{$x=(x_0,x_1,x_2,x_3)$}. 

$\Pi_{\mu\nu}(x;y)$ may be obtained as the correlation function for two
current operators due to the charge carriers of the media: 
\begin{equation}\label{eq:defpi}
	\Pi_{\mu\nu}(x;y)\;=\; \langle j_\mu(x)
	j_\nu(y) \rangle \;=\; \Pi_{\nu\mu}(y;x)\;,
\end{equation}
and, since those currents are localized on the mirrors, $\Pi_{\mu\nu}(x;y)$
will be different from zero only when $x_3$ and $y_3$ are inside the media.
Namely, when they belong to a certain interval containing $0$, the center
of the mirror.

The current operator is assumed to be conserved, so that ${\Pi}_{\mu
\nu}(x,y)$ satisfies the Ward identity:
\begin{equation}\label{eq:Ward}
	\partial^x_{\mu}\Pi_{\mu \nu}(x;y) \;=\; 0 \;,
\end{equation}
where $\mu,\nu=0,1,2,3$, and $\partial^x_\mu \equiv \frac{\partial}{\partial
x_\mu}$.

Symmetry under translations along the `parallel' spacetime coordinates
\mbox{$x_\parallel \equiv (x_0,x_1,x_2)$}, 
\mbox{$y_\parallel \equiv (x_0,x_1,x_2)$}, 
implies that
\begin{equation}\label{eq:psym}
	\Pi_{\mu \nu}(x;y) \;=\; \Pi_{\mu\nu}(x_\parallel,
	x_3;y_\parallel, y_3) \;=\; \Pi_{\mu
	\nu}(x_\parallel-y_\parallel;x_3,y_3)\;.
\end{equation}
To proceed, we take advantage of (\ref{eq:psym}) to rephrase
(\ref{eq:Ward}) in terms of the  Fourier transform of ${\Pi}_{\mu\nu}$ with
respect to $x_\parallel-y_\parallel$,
\begin{equation}\label{eq:pitrans}
	\Pi_{\mu \nu}(x_\parallel-y_\parallel;x_3,y_3) \;=\;
\int \frac{d^3k_\parallel}{(2\pi)^3} \; 
e^{i k_\parallel \cdot (x_\parallel-y_\parallel)}
\, \widetilde{\Pi}_{\mu \nu}(k_\parallel;x_3,y_3) \;.
\end{equation}
Reality and Bose symmetry of the currents implies that
$\widetilde{\Pi}_{\mu \nu}$ satisfies the identities:
\begin{equation}\label{eq:props}
\left\{ \begin{array}{ccc}
		\widetilde{\Pi}_{\mu \nu}(k_\parallel;x_3,y_3) &=&
		\widetilde{\Pi}^*_{\mu \nu}(-k_\parallel;x_3,y_3)\\
		\widetilde{\Pi}_{\mu \nu}(k_\parallel;x_3,y_3)  &=& 
		\widetilde{\Pi}_{\nu\mu}(-k_\parallel;y_3,x_3) \;\;\;.
\end{array}
\right.
\end{equation}
Then we write (\ref{eq:Ward}) in Fourier space; to do so,
it is convenient to use different indices for the translation invariant
directions than for $x_3$. To that end, we adopt the convention that indices from
the beginning of the Greek alphabet: \mbox{$\alpha,\beta,\ldots$} run from
$0$ to $2$. Then: 
\begin{eqnarray}\label{eq:Ward3}
	i k_{\alpha}\widetilde{\Pi}_{\alpha \beta}(k_\parallel;x_3,y_3) \;+\;
	\partial_3 \widetilde{\Pi}_{3 \beta}(k_\parallel;x_3,y_3)&=& 0 \nonumber\\
	i k_{\alpha}\widetilde{\Pi}_{\alpha 3}(k_\parallel;x_3,y_3) \;+\;
	\partial_3 \widetilde{\Pi}_{3 3}(k_\parallel;x_3,y_3)&=& 0 \;\;\;.
\end{eqnarray}

We can then use again the (assumed) symmetries, to write the different components of
$\widetilde{\Pi}_{\mu\nu}$ in simpler terms, in particular those appearing
in the Ward identity. For example, regarding the $\alpha,\beta$ components, 
we see that $\widetilde{\Pi}_{\alpha\beta}$  can be decomposed as follows:
\begin{equation}\label{eq:decpiab}
	\widetilde{\Pi}_{\alpha \beta}(k_\parallel;x_3,y_3)\;=\;
	\widetilde{\Pi}^\perp_{\alpha \beta}(k_\parallel;x_3,y_3) \,+\,
	B(|k_\parallel|;x_3,y_3) \; {\mathcal Q}_{\alpha \beta}(k_\parallel)
\end{equation}
where $\widetilde{\Pi}^\perp_{\alpha \beta}$ is $2+1$ dimensional transverse:
\begin{equation}
k_\alpha \widetilde{\Pi}^\perp_{\alpha \beta}(k_\parallel;x_3,y_3)\;=\; 0 \;,
\end{equation}
$B$ is a $2+1$ dimensional scalar, and ${\mathcal Q}$ is the longitudinal
proyector,
\begin{equation}\label{eq:defproyq}
{\mathcal Q}_{\alpha \beta}(k_\parallel) \;=\;
\frac{k_{\alpha}k_{\beta}}{k_\parallel^2} \;.
\end{equation}
Besides, $B$ is real and symmetric under the interchange of $x_3$ with
$y_3$. Note that $B$, but not $\widetilde{\Pi}_{\alpha\beta}^\perp$,  enter in the Ward identity.

We may also write 
\begin{equation}\label{eq:decpia3}
\widetilde{\Pi}_{\alpha 3}(k_\parallel;x_3,y_3) \;=\; i \, k_\alpha \,
C(|k_\parallel|;x_3,y_3) \;,
\end{equation}
in terms of a real and scalar function $C(|k_\parallel|;x_3,y_3)$.
Using the properties in (\ref{eq:props}), we see that:
\begin{equation}\label{eq:decpi3a}
\widetilde{\Pi}_{3 \alpha}(k_\parallel;x_3,y_3) \;=\; - i \, k_\alpha \,
C(|k_\parallel|;y_3,x_3) \;.
\end{equation}

Inserting the decompositions (\ref{eq:decpiab}) and (\ref{eq:decpi3a}) into
(\ref{eq:Ward3}), we obtain the two independent relations:
\begin{eqnarray}\label{eq:Ward4}
B(k_\parallel;x_3,y_3) &=& - \frac{\partial C}{\partial x_3}(k_\parallel;y_3,x_3) \nonumber\\
C(k_\parallel;x_3,y_3) &=& \frac{1}{k_\parallel^2} 
	\frac{\partial \widetilde{\Pi}_{33}}{\partial x_3}(k_\parallel;x_3,y_3)\;.
\end{eqnarray}
These two relations then allow us to determine $B$ and $C$ in terms of the
single real and symmetric function $\widetilde{\Pi}_{33}$. 
Then, any  given model for the mirror will be characterized by a choice of
$\widetilde{\Pi}^\perp_{\alpha\beta}$ and
$\widetilde{\Pi}_{33}$; the rest becomes determined, since:
\begin{eqnarray}\label{eq:Ward5}
B(k_\parallel;x_3,y_3) &=& - \frac{1}{k_\parallel^2} 
\frac{\partial^2 \widetilde{\Pi}_{33}}{\partial x_3 \partial y_3}(k_\parallel;x_3,y_3) \nonumber\\
C(k_\parallel;x_3,y_3) &=& \frac{1}{k_\parallel^2} 
\frac{\partial \widetilde{\Pi}_{33}}{\partial x_3}(k_\parallel;x_3,y_3)\;.
\end{eqnarray}
Thus the expanded expression for ${\mathcal S}_M$ is: 
\begin{eqnarray}\label{eq:sides}
S_M=\frac{1}{2} \int_{k_{\parallel};x_3,y_3} \tilde{A}^*_{\mu}(k_{\parallel},x_3)
\widetilde{\Pi}_{\mu \nu}(k_\parallel;x_3,y_3)\tilde{A}_{\nu}(k_{\parallel},y_3)\nonumber\\
= \frac{1}{2} \int_{k_{\parallel};x_3,y_3} \tilde{A}^*_{\alpha}(k_{\parallel},x_3)
\widetilde{\Pi}_{\alpha \beta}(k_\parallel;x_3,y_3)\tilde{A}_{\beta}(k_{\parallel},y_3)\nonumber\\
+\frac{1}{2} \int_{k_{\parallel};x_3,y_3} \tilde{A}^*_{3}(k_{\parallel},x_3)
\widetilde{\Pi}_{3 3}(k_\parallel;x_3,y_3)\tilde{A}_{3}(k_{\parallel},y_3)\nonumber\\
+\frac{1}{2} \int_{k_{\parallel};x_3,y_3} \tilde{A}^*_{3}(k_{\parallel},x_3)
\widetilde{\Pi}_{3 \alpha}(k_\parallel;x_3,y_3)\tilde{A}_\alpha(k_{\parallel},y_3)\nonumber\\
+\frac{1}{2} \int_{k_{\parallel};x_3,y_3} \tilde{A}^*_{\alpha}(k_{\parallel},x_3)
\widetilde{\Pi}_{\alpha 3}(k_\parallel;x_3,y_3)\tilde{A}_{3}(k_{\parallel},y_3)\;,
\end{eqnarray}
and may be written, taking into account our previous discussion, as follows,
\begin{eqnarray}\label{eq:sides1}
{\mathcal S}_M &=& \frac{1}{2} \int_{k_{\parallel};x_3,y_3} \tilde{A}^*_{\alpha}(k_{\parallel},x_3)
\widetilde{\Pi}^\perp_{\alpha \beta}(k_\parallel;x_3,y_3)\tilde{A}_{\beta}(k_{\parallel},y_3)\nonumber\\
&-&\frac{1}{2} \int_{k_{\parallel};x_3,y_3} 
\tilde{A}^*_\alpha(k_{\parallel},x_3)
\,\frac{\partial^2 \widetilde{\Pi}_{33}}{\partial x_3 \partial y_3}(k_\parallel,x_3,y_3)  
\frac{k_\alpha k_\beta}{k_\parallel^4} \tilde{A}_{\beta}(k_{\parallel},y_3)\nonumber\\
&+&\frac{1}{2} \int_{k_{\parallel};x_3,y_3} \tilde{A}^*_{3}(k_{\parallel},x_3)
\widetilde{\Pi}_{3 3}(k_\parallel;x_3,y_3)\tilde{A}_{3}(k_{\parallel},y_3)\nonumber\\
&-&\frac{i}{2} \int_{k_{\parallel};x_3,y_3} 
\tilde{A}^*_{3}(k_{\parallel},x_3)
\frac{k_\alpha}{k_\parallel^2} \frac{\partial \widetilde{\Pi}_{3 3}}{\partial y_3}(k_\parallel;y_3,x_3)
\tilde{A}_{\alpha}(k_{\parallel},y_3)\nonumber\\
&+&\frac{i}{2} \int_{k_{\parallel};x_3,y_3} \tilde{A}^*_{\alpha}(k_{\parallel},x_3)
\frac{k_\alpha}{k_\parallel^2} \frac{\partial \widetilde{\Pi}_{3 3}}{\partial x_3}(k_\parallel;x_3,y_3)
\tilde{A}_{3}(k_{\parallel},y_3) \;\;\;.
\end{eqnarray}

The interaction terms ${\mathcal S}_L$ and ${\mathcal S}_R$ will then have the form:
\begin{equation}
	{\mathcal S}_L (A) \;=\; \frac{1}{2} \int
	\frac{d^3k_\parallel}{(2\pi)^3} \int_{x_3,y_3}
	\tilde{A}^*_\mu(k_\parallel,x_3)
	\widetilde{\Pi}^{(L)}_{\mu\nu}(k_\parallel;x_3,y_3)\tilde{A}_\nu(k_\parallel,y_3)\;,
\end{equation}
and
\begin{equation}
	{\mathcal S}_R (A) \;=\; \frac{1}{2} \int
	\frac{d^3k_\parallel}{(2\pi)^3} \int_{x_3,y_3}
	\tilde{A}^*_\mu(k_\parallel,x_3)
	\widetilde{\Pi}^{(R)}_{\mu\nu}(k_\parallel;x_3,y_3)\tilde{A}_\nu(k_\parallel,y_3)\;,
\end{equation}
where $\widetilde{\Pi}^{(L)}_{\mu\nu}$ and $\widetilde{\Pi}^{(R)}_{\mu\nu}$
contain (independent) functions $B^{(L,R)}$ and $C^{(L,R)}$, determined by
the respective $\widetilde{\Pi}^{(L,R)}_{33}$ components, which are
concentrated around $x_3=0$ and $x_3=a$, respectively.

We also need the full action ${\mathcal S}$ in the `mixed' Fourier
representation we have just introduced for the interaction term. In this
form, the free part ${\mathcal S}_0$ is given by:
\begin{eqnarray}\label{eq:freetermFourier}
	S_0 &=&\frac{1}{2} \int_{{\mathbf k_{\parallel}},x_3} \Big\{
\tilde{A}^*_{\alpha} \big[(-\partial_3^2 + k_\parallel^2 )\delta_{\alpha \beta}
- k_\alpha k_\beta \big]\tilde{A}_{\beta} 
\,+\, \tilde{A}^*_3 \,k^2_{\parallel}\, \tilde{A}_3\nonumber\\
&+& (i k_{\alpha} \partial_3 \tilde{A}_\alpha)^* \tilde{A}_3
+ \tilde{A}^*_3 ( i k_{\alpha} \partial_3
\tilde{A}_{\alpha}) \Big\} \;.
\end{eqnarray}

\section{Reduction to one-dimensional systems}\label{sec:reduction}
Each mirror has been characterized by its vacuum polarization tensor
$\widetilde{\Pi}^{(L)}$, $\widetilde{\Pi}^{(R)}$. 
Before proceeding, we introduce a convenient gauge fixing. It is
not difficult to realize that, for the system at hand, a highly convenient choice is:
\begin{equation}\label{eq:gaugefixing}
	\partial_\alpha A_\alpha (x) \;=\;0 \;, 
\end{equation}
which leads to a gauge-fixed action which may be split as follows:
\begin{equation}\label{eq:sdecomp}
	{\mathcal S}(A) \;=\; {\mathcal S}_\parallel(A_\parallel) \,+\, 
	{\mathcal S}_3(A_3)  \;,
\end{equation}
with:
\begin{eqnarray}\label{eq:sparallel}
{\mathcal S}_\parallel (A_\parallel) &=& \frac{1}{2}
\int_{k_{\parallel};x_3,y_3} \Big\{
\tilde{A}^*_{\alpha}(k_{\parallel},x_3)\big[
(-\partial^2_{x_3} + k_\parallel^2)\delta(x_3-y_3) \big]
{\mathcal P}^\perp_{\alpha \beta} \tilde{A}_{\beta}(k_{\parallel},y_3) \nonumber\\
&+&\tilde{A}^*_{\alpha}(k_{\parallel},x_3) \widetilde{\Pi}^{(L)\perp}_{\alpha
\beta}(k_\parallel;x_3,y_3) \tilde{A}_{\beta}(k_{\parallel},y_3)\nonumber\\
&+&\tilde{A}^*_{\alpha}(k_{\parallel},x_3) \widetilde{\Pi}^{(R)\perp}_{\alpha \beta}(k_\parallel;x_3,y_3)
\tilde{A}_{\beta}(k_{\parallel},y_3) \Big\} \;,
\end{eqnarray}
where ${\mathcal P}^{\perp}_{\alpha \beta} \equiv \delta_{\alpha \beta} -
\frac{k_{\alpha}k_{\beta}}{k_\parallel^2}$,  and
\begin{eqnarray}\label{eq:s3}
	{\mathcal S}_3 &=& \frac{1}{2} \int_{k_{\parallel};x_3,y_3} \Big[
		\tilde{A}^*_3(k_{\parallel},x_3) \, k^2_{\parallel}\,
		\delta(x_3-y_3) \tilde{A}_3(k_{\parallel},y_3) \nonumber\\
		&+& \tilde{A}^*_{3}(k_{\parallel},x_3) \widetilde{\Pi}^{(L)}_{33}(k_\parallel;x_3,y_3)
\tilde{A}_{3}(k_{\parallel},y_3) \nonumber\\
&+& \tilde{A}^*_{3}(k_{\parallel},x_3) \widetilde{\Pi}^{(R)}_{3 3}(k_\parallel;x_3,y_3)
\tilde{A}_{3}(k_{\parallel},y_3) \Big] \;.
\end{eqnarray}
When evaluating the functional integral over the (gauge-fixed) gauge field
configurations, we see that, since the action is decomposed as the sum of
an $A_\alpha$ plus an $A_3$ term, with no mixing,  
\begin{equation}\label{eq:defzeta}
\frac{{\mathcal Z}}{{\mathcal Z}_0} \,=\, 
\frac{{\mathcal Z}_\parallel}{{\mathcal Z}_{\parallel 0}}
\times  
\frac{{\mathcal Z}_3}{{\mathcal Z}_{3, 0}} 
\end{equation}
with:
\begin{equation}\label{eq:defzetapar3}
	{\mathcal Z}_\parallel \;=\; \int \big[{\mathcal
	D}\tilde{A}_\parallel\big] \, e^{-{\mathcal S}_\parallel({\tilde
	A}_\parallel)} \;,\;\;
	{\mathcal Z}_3 \;=\; \int {\mathcal D}\tilde{A}_3 \, 
	e^{-{\mathcal S}_3({\tilde A}_3)} \;.
\end{equation}

To proceed, we note that each  $\widetilde{\Pi}^\perp_{\alpha \beta}$
may be decomposed into two irreducible transverse tensors (projectors), in
terms of two scalars. Indeed, the assumed isotropy and homogeneity of the
media along the parallel directions, means that we can construct two
independent transverse tensors using as building blocks the elements:
$\breve{k}_{\alpha}\equiv k_{\alpha} - k_0 n_{\alpha}$, and
$\breve{\delta}_{\alpha \beta} \equiv \delta_{\alpha \beta} -
n_{\alpha}n_{\beta}$, where $n=(1,0,0)$. 

Two independent projectors ${\mathcal P}^{t}$ and ${\mathcal P}^{l}$ that are
transverse to $k_\alpha$ may be written as follows:
\begin{equation}
	{\mathcal P}^{t}_{\alpha \beta} \equiv {\breve{\delta}_{\alpha \beta}} - \frac
{\breve{k}_{\alpha} \breve{k}_{\beta}}{{\breve{k}}^2}
\end{equation}
and
\begin{equation}
	{\mathcal P}^{l}_{\alpha \beta} \equiv {\mathcal P}^{\perp}_{\alpha
	\beta} - {\mathcal P}^{t}_{\alpha \beta} \;.
\end{equation}
They satisfy the following algebraic properties:
	$$
	{\mathcal P}^\perp + {\mathcal Q} = I \;,\;\;
	{\mathcal P}^t + {\mathcal P}^l = {\mathcal P}^\perp 
	$$
	$$
	{\mathcal P}^t {\mathcal P}^l = {\mathcal P}^l {\mathcal P}^t = 0
	\;,\;\; {\mathcal Q} {\mathcal P}^t = {\mathcal P}^t {\mathcal Q} = 0 
	\;,\;\; {\mathcal Q} {\mathcal P}^l = {\mathcal P}^l {\mathcal Q} = 0 
	$$
\begin{equation}
\big({\mathcal P}^\perp\big)^2 = {\mathcal P}^\perp \;,
\big({\mathcal Q})^2 = {\mathcal Q} \;,
\big({\mathcal P}^t\big)^2 = {\mathcal P}^t \;,
\big({\mathcal P}^l\big)^2 = {\mathcal P}^l \;,
\end{equation}
where $I_{\alpha\beta} = \delta_{\alpha\beta}$.
Therefore we can express $\widetilde{\Pi}_{\alpha \beta}^{\perp(I)}$ $(I=L,R)$, as follows:
\begin{equation}\label{polarizationdelvacio}
\tilde{\Pi}_{\alpha \beta}^{\perp(I)}(k_\parallel; x_3,y_3) \;=\; f^{(I)}_t ({k}^{2}_{0},
\mathbf{k}_\parallel^{2}; x_3,y_3) \, 
{\mathcal P}^{t}_{\alpha \beta} +
f^{(I)}_l(k^{2}_{0};\mathbf{k}_\parallel^{2}, x_3,y_3)
{\mathcal P}^{l}_{\alpha \beta} \;.	
\end{equation}

A more explicit form for ${\mathcal S}_\parallel$ is then obtained
\begin{eqnarray}
	{\mathcal S}_\parallel = \frac{1}{2} \int_{k_{\parallel};x_3,y_3} 
	\widetilde{A}^*_\alpha(k_\parallel,x_3) \Big[\widetilde{K}_{t}(k_\parallel) {\mathcal P}^{t}_{\alpha \beta}
		+  \widetilde{K}_{l}(k_\parallel) {\mathcal P}^{l}_{\alpha \beta} \Big]
\widetilde{A}_\beta(k_\parallel,x_3) \;,
\end{eqnarray}
where
\begin{eqnarray}
	\widetilde{K}_{t,l}(k_\parallel;x_3,y_3) &=& (-\partial_3^2 +
	k_\parallel^2) \delta(x_3-y_3) +
	\widetilde{V}_{t,l}(k_\parallel;x_3,y_3) \;,
\end{eqnarray}	
and:	
\begin{equation}	
	\widetilde{V}_{t,l}(k_\parallel;x_3,y_3) \,=\, \sum_{I=L,R} f^{(I)}_{t,l} ({k}^{2}_{0}, 
	\mathbf{k}_\parallel^{2}; x_3, y_3)  \;,
\end{equation}
what concludes the reduction. Indeed, note that the action has been reduced
to a quadratic form  for an operator which has been decomposed into orthogonal
rank-one projectors.

\section{Casimir energy}\label{sec:energy}
Because of (\ref{eq:defzeta}), the total Casimir energy shall be a sum
of a contribution coming from the integral over $A_3$, plus another from
$A_\parallel$.
Let us consider first the $A_3$ one. Defining:
\begin{equation}
	\Gamma_3 \,= - \, \lim_{L\to \infty} 
	\Big[ \frac{1}{L^2} \, \log \frac{{\mathcal
	Z}_3}{{\mathcal Z}_{3,0}}\Big] \;,
\end{equation}
we see that it does not contain any contribution to the {\em interaction\/}
energy between the mirrors. Indeed, even for the case of generally nonlocal
$\widetilde{\Pi}^{(L,R)}_{33}$ functions, if their arguments have
non-overlapping support (for the different mirrors), $\Gamma_3$ is
independent of the distance between the mirrors. The reason is that the
$A_3$ field does not propagate in the normal direction, in the sense that
its free action is ultralocal in this gauge:
\begin{equation}
	{\mathcal S}_3(A_3) \,=\,  \frac{1}{2} \int_{k_{\parallel};x_3,y_3}
		\tilde{A}^*_3(k_{\parallel},x_3) \, k^2_{\parallel}\,
		\delta(x_3-y_3) \tilde{A}_3(k_{\parallel},y_3) \;.
\end{equation}
Regarding the $A_\parallel$ contribution, using properties of the
projectors, we see that the effective action becomes:
\begin{equation}
	\Gamma_C \,=\,\Gamma_t + \Gamma_l  
\end{equation}
where
\begin{equation}
	\Gamma_{t,l} \,= - \, \lim_{L\to \infty} 
	\Big[ \frac{1}{L^2} \, \log \frac{{\mathcal
	Z}^{(t,l)}}{{\mathcal Z}^{(t,l)}_0}\Big]
\end{equation}
or
\begin{equation}
	\Gamma_{t,l} \,= \, \frac{1}{2} \,\int
	\frac{d^3k_{\parallel}}{(2 \pi)^3}\,
	\log \Big[\frac{\det \widetilde{K}_{t,l}(k_\parallel)}{\det \widetilde{K}_0(k_\parallel)}\Big] \;
\end{equation}
where $\widetilde{K}_0=\widetilde{K}_{t,l}\lvert_{V=0}$.

The system has been reduced to two independent Casimir problems, each one
of them corresponding to a real scalar field in the presence of a generally
nonlocal potential background $\widetilde{V}_{t,l}$. This is the main result 
of this letter, namely, the fact that if the EM properties of the mirrors proceed 
from charges with a conserved current, the system can be reduced to two independent scalar-like
problems.

Local potential
backgrounds have been extensively studied in~\cite{MIT}, in a Quantum Field
Theory set up, within the context of the Dirichlet Casimir problem for a
real scalar field. Here, we shall, as a final step in the calculation, express the result for
the Casimir energy for those decoupled systems in terms of a
Lifshitz formula~\cite{Lifshitz}. 
To that end, we shall use the approach developed
in~\cite{CcapaTtira:2011ga}, noting that the potentials are built in terms of the functions that appear in the
decomposition of the transverse part of the vacuum polarization tensor into a set of irreducible tensors.
For the case of $x_3$-local functions:
\begin{eqnarray}
f^{(L)}_{t,l} ({k}^{2}_{0}, \mathbf{k}_\parallel^{2}; x_3, y_3)  &=&
f^{(L)}_{t,l} ({k}^{2}_{0}, \mathbf{k}_\parallel^{2}; x_3) \delta(x_3-y_3)
\nonumber\\
f^{(R)}_{t,l} ({k}^{2}_{0}, \mathbf{k}_\parallel^{2}; x_3, y_3)  &=&
f^{(R)}_{t,l} ({k}^{2}_{0}, \mathbf{k}_\parallel^{2}; x_3 - a)
\delta(x_3-y_3) \;,
\end{eqnarray}
where each function is localized around $x_3 =0$. In this case, we may
apply the general formula derived in~\cite{CcapaTtira:2011ga}, to write: 
\begin{equation}\label{eq:resgtl}
	\Gamma_{t,l} \,= \, \frac{1}{2} \, \int_{k_\parallel} 
	\log \left[ 1 + \frac{T^{(R)}_{12}}{T^{(L)}_{11}}
		\frac{T^{(R)}_{21}}{T^{(L)}_{11}} e^{- 2\,
	\mid k_\parallel\mid \, l} \right]_{t,l}\;,
\end{equation}
where $T_{t,l}$ is the result of performing the following change of basis to the matrix $A_{t,l}$:
\begin{equation}
T_{t,l}= B^{-1}A_{t,l}B
\end{equation}
with
\begin{equation}
B=\frac{1}{\sqrt2}
\left( \begin{array}{rr}
1 & 1 \\
1 & -1
\end{array} \right),
\end{equation}
and $A_{t,l}$ are defined as in~\cite{CcapaTtira:2011ga},
regarding each one, $t$ or $l$, as due to an independent field, in its own
background potential.

Thus, in the case of two finite width mirrors represented by local potential
backgrounds, the Casimir force between them can be calculated applying the Lifshitz formula
to two independent scalar-like problems.

\section{Conclusions}\label{sec:conclusions}
We have shown that the Casimir interaction energy corresponding to two
flat and parallel finite width mirrors coupled to the EM field can be
decomposed into two contributions, where each one of them can be evaluated
as in an equivalent scalar field problem. We have shown that under the
assumption of translation invariance along the parallel coordinates, but
without imposing extra conditions on the dependence of the mirrors'
properties along the normal coordinates, which may even by nonlocal~\cite{Fosco:2009cw}.
For the local case, the problem has been treated by the approach
of~\cite{CcapaTtira:2011ga}, which was based in turn on~\cite{GY,reviewGY}.

The Ward identity for the vacuum polarization tensor corresponding to each
mirror plays an important role in the decoupling of the system into two
separated independent problems. This phenomenon manifests itself
more clearly in a particular gauge, however, it may be checked in other
cases, as well, albeit at the expense of a rather involved calculation,
since $A_3$ becomes entangled to the other components of the gauge field.
Nevertheless, we have explicitly checked that to be true in other gauges,
like the Feynman gauge, and the axial ($A_3=0$) gauge.

\section*{Acknowledgements}
This work was supported by ANPCyT, CONICET, and UNCuyo.


\begin{thebibliography}{bib}
\bibitem{Casimir}H.~B.~G.~Casimir,
%``On the Attraction Between Two Perfectly Conducting Plates,''
Indag.\ Math.\  {\bf 10}, 261 (1948)
[Kon.\ Ned.\ Akad.\ Wetensch.\ Proc.\  {\bf 51}, 793 (1948)]
[Front.\ Phys.\  {\bf 65}, 342 (1987)]
[Kon.\ Ned.\ Akad.\ Wetensch.\ Proc.\  {\bf 100N3-4}, 61 (1997)].
\bibitem{rev}G. Plunien, B. M\"uller, and W. Greiner, Phys. Rep.
	\textbf{134}, 87 (1986); P. Milonni, {\it The Quantum Vacuum}
	(Academic Press, San Diego, 1994); V. M. Mostepanenko and N. N.
	Trunov, {\it The Casimir Effect and its Applications} (Clarendon,
	London, 1997); M.  Bordag, {\it The Casimir Effect 50 Years Later}
	(World Scientific, Singapore, 1999); M. Bordag, U. Mohideen, and V.
	M. Mostepanenko, Phys. Rep.  \textbf{353}, 1 (2001); K. A. Milton,
	{\it The Casimir Effect: Physical Manifestations of the Zero-Point
	Energy} (World Scientific, Singapore, 2001); S. Reynaud {\it et
	al.}, C. R. Acad.  Sci. Paris \textbf{IV-2}, 1287 (2001); K. A.
	Milton, J. Phys. A: Math. Gen. \textbf{37}, R209 (2004); S.K.
	Lamoreaux, Rep. Prog.  Phys. \textbf{68}, 201 (2005); Special Issue
	{\it "Focus on Casimir Forces"}, New J. Phys. \textbf {8} (2006).  
\bibitem{Fosco:2008td} 
  C.~D.~Fosco, F.~C.~Lombardo and F.~D.~Mazzitelli,
  %``Casimir effect with dynamical matter on thin mirrors,''
  Phys.\ Lett.\ B {\bf 669}, 371 (2008)
  [arXiv:0807.3539 [hep-th]].
\bibitem{MIT} N. Graham, R. L. Jaffe,  V. Khemani,  M. Quandt,  O.  Schroeder, and  H. Weigel,
Nucl. Phys. B {\bf 677}, 379 (2004). 
\bibitem{Lifshitz} E. M. Lifshitz, Sov. Phys. JETP {\bf 2}, 73 (1956).
%\cite{CcapaTtira:2011ga}
\bibitem{CcapaTtira:2011ga} 
C.~Ccapa Ttira, C.~D.~Fosco and F.~D.~Mazzitelli,
%``Lifshitz formula for the Casimir force and the Gelfand-Yaglom theorem,''
J.\ Phys.\ A A {\bf 44}, 465403 (2011).
\bibitem{Fosco:2009cw}C.~D.~Fosco and E.~Losada,
%``Casimir effect with nonlocal boundary interactions,''
Phys.\ Lett.\ B {\bf 675}, 252 (2009)
[arXiv:0902.2198 [hep-th]].
\bibitem{GY}I.~M.~Gelfand and A.~M.~Yaglom,
%``Integration in functional spaces and it applications in quantum physics,''
J.\ Math.\ Phys.\  {\bf 1}, 48 (1960).
\bibitem{reviewGY}G.~V.~ Dunne, J. Phys. A {\bf 41}, 304006 (2008).
\end{thebibliography}
\end{document}